\documentclass[%
 reprint,
superscriptaddress,
nofootinbib,
 amsmath,amssymb,
 aps,
]{revtex4-2}

\usepackage[colorlinks,linktocpage,linkcolor=cyan,citecolor=cyan]{hyperref}

\usepackage{graphicx}
\usepackage{dcolumn}
\usepackage{bm}
\usepackage{hyperref}

\def\MP{M_{\rm Pl}}

\begin{document}
\begin{flushleft} {\footnotesize IPMU22-0044, YITP-22-93} \end{flushleft}

\author{Misao Sasaki}
\email{misao.sasaki@ipmu.jp}
\affiliation{Kavli Institute for the Physics and Mathematics of the Universe (WPI), UTIAS, The University of Tokyo, Chiba 277-8583, Japan}
\affiliation{Center for Gravitational Physics and Quantum Information, Yukawa Institute for Theoretical Physics, Kyoto University, Kyoto 606-8502, Japan}
\affiliation{Leung Center for Cosmology and Particle Astrophysics, National Taiwan University, Taipei 10617, Taiwan}

\author{Valeri Vardanyan}
\email{valeri.vardanyan@ipmu.jp}
\affiliation{Kavli Institute for the Physics and Mathematics of the Universe (WPI), UTIAS, The University of Tokyo, Chiba 277-8583, Japan}

\author{Vicharit Yingcharoenrat}
\email{vicharit.yingcharoenrat@ipmu.jp }
\affiliation{Kavli Institute for the Physics and Mathematics of the Universe (WPI), UTIAS, The University of Tokyo, Chiba 277-8583, Japan}

\title{Super-horizon resonant magnetogenesis during inflation}

\begin{abstract}
We propose a novel mechanism for significantly enhancing the amplitude of primordial electromagnetic fields during inflation. Similar to existing proposals, our idea is based on parametric resonance effects due to conformal-symmetry-breaking coupling of a gauge field and the inflaton. Our proposed scenario, however, significantly differs from previously studied models, and avoids their shortcomings. We, particularly, construct a viable system where the gauge field is exponentially amplified on super-horizon scales, therefore evading the no-go theorem formulated on the basis of widely encountered drastic back-reaction of the magnetic field energy on the inflationary background. We argue that in order for the resonant scenario to work with a bounded and positive-definite coupling function, a mass term for the gauge sector is required. We compute the spectrum of the produced magnetic fields and demonstrate the compatibility with current observational constraints. We demonstrate that while the magnetic fields do not noticeably back-react on the inflationary background, the non-zero mass term can contribute significantly to the total energy-momentum tensor. We point out the parameter space where the latter issue is absent.

\end{abstract}

\maketitle


\section{Introduction}

The origin of cosmological magnetic fields on intergalactic scales of $\sim 4~{\rm Mpc}$ remains unexplained in cosmology. The strength of such magnetic fields has been measured to be between $10^{-17}$ and $10^{-14}$ Gauss by several experiments \cite{Ando:2010rb,Tavecchio_2010,Neronov_2010,Tavecchio_2011,Essey:2010nd,Finke:2015ona}. Interestingly, inflation -- an accelerated expansion in the early universe -- may be a working regime where the quantum vacuum fluctuations of magnetic fields can be stretched beyond the horizon and seed the observed magnetic fields on large scales. There are many models proposed in the literature \cite{PhysRevD.37.2743,1992ApJ...391L...1R,PhysRevD.46.5346,Dolgov:1993vg,Gasperini:1995dh,Martin:2007ue,Demozzi:2009fu,Kanno:2009ei,Emami:2009vd,Bamba:2003av,Bamba:2006ga,Kobayashi:2014sga,Barnaby:2012tk,BazrafshanMoghaddam:2017zgx,Durrer:2022emo,Kushwaha:2020nfa}, most of which are based on breaking of the conformal invariance in the gauge sector (see also Refs.~\cite{Kandus:2010nw,Subramanian:2009fu} for useful reviews). This is necessary because in the Maxwell's theory there is no enhancement of the electromagnetic (EM) fluctuations in a Friedmann–Lemaitre-Robertson-Walker (FLRW) universe.

There are multiple ways for breaking the conformal invariance. The simplest one is to introduce a coupling of the EM field and a (pseudo) scalar inflaton, although alternatives involving coupling the gauge field to the scalar curvature \cite{Mazzitelli:1995mp,PhysRevD.37.2743} or additional spectator fields \cite{Giovannini:2007rh,Patel:2019isj,Giovannini:2021thf} have also been considered. In fact, one of the well-studied models of magnetogenesis during inflation is the so-called Ratra model (see e.g. Refs.~\cite{Talebian:2021dfq,Talebian:2020drj,BazrafshanMoghaddam:2017zgx} for recent works) with a non-minimal coupling of the form $f(\phi)^2 F_{\mu\nu}F^{\mu\nu}$, where $\phi$ is the inflaton field and $F_{\mu\nu}$ is the electromagnetic field strength. Typically, there are two problems in such models of primordial magnetogenesis; these are the strong-coupling \cite{Fujita:2012rb,Ferreira:2013sqa,Ferreira:2014hma,BazrafshanMoghaddam:2017zgx} and back-reaction \cite{Demozzi:2009fu,Kanno:2009ei,Fujita:2012rb,Green:2015fss,Fujita:2016qab} problems. The former arises when the effective coupling constant ($\sim f(\phi)^{-1}$) becomes much larger compared to unity, so that the perturbative calculations are unreliable. The latter problem usually happens when the energy density of the electromagnetic sector becomes comparable to the background energy density during inflation, prematurely ending it. Additionally, there are phenomenological limitations arising from Cosmic Microwave Background (CMB) constraints \cite{Barnaby:2012tk,Giovannini:2013rme,Fujita:2013qxa,Fujita:2016qab}. 

Among the interesting models of primordial magnetogenesis, the idea of resonant production during inflation was investigated in Refs.~\cite{Byrnes:2011aa,Patel:2019isj}. In Ref.~\cite{Byrnes:2011aa}, a coupling of the form $I(\phi) F_{\mu\nu}\widetilde{F}^{\mu\nu}$ was considered, where $I(\phi)$ is a function of the inflaton $\phi$ and $\widetilde{F}_{\mu\nu}$ is the dual of the EM field strength. Assuming that $I(\phi)$ is an oscillating function in conformal time, the helical EM modes with sub-horizon wavelengths grow exponentially, sourcing the late-universe large-scale magnetic fields. However, since the amplification is predominantly at sub-horizon scales during inflation, the magnetic energy density dilutes rapidly when the amplified modes exit the horizon. In order for the heavily diluted magnetic energy density to be in agreement with observational constraints, the sub-horizon enhancement mechanism should be very powerful, leading to magnetic field energy density at the epoch of horizon-crossing exceeding the inflationary background energy density (see Ref.~\cite{Byrnes:2011aa} as well as Section~\ref{sec:no-go} below). 

In this work, we consider a coupling of the form $f(\phi)^2 F_{\mu\nu}F^{\mu\nu}$ together with the mass term of $A_\mu$ in the context of a resonant mechanism for generating a sizable primordial magnetic field. Oscillations of $f(\phi)$ (hence in the coupling of the EM sector) and a small mass of $A_\mu$ compared to Hubble scale can give rise to resonant amplification of certain EM modes, without leading to both the strong-coupling and back-reaction problems. We will see in Section~\ref{sec:toy_model} that the no-go theorem mentioned above does not apply to our model due to the fact that the significant enhancement of the EM modes takes place outside the horizon. We will argue that in order for the resonant scenario to work with a bounded and positive-definite coupling function $f(\phi)$ a mass term for the gauge sector is required. We will demonstrate that while the magnetic fields do not noticeably back-react on the inflationary background, this mass term contributes significantly to the total energy-momentum tensor. We point out the parameter space where the latter issue is absent.

The rest of this paper is organized as follows. In Section~\ref{sec:setup} we introduce our model along with the notation used throughout the paper. We summarize the no-go argument discussed in Ref.~\cite{Byrnes:2011aa} and explain a possible way out in Section~\ref{sec:no-go}. In Section~\ref{sec:toy_model} we analyze a toy model which generates large late-time magnetic fields without encountering both the strong-coupling and back-reaction problems. Our conclusions are summarized in Section~\ref{sec:conclusion}, and technical details presented in Appendix~\ref{app:Hill_equation}.

\section{Setup}\label{sec:setup}
The model considered here is given by the following action,
\begin{align}
    S = \int \mathrm{d}^4x \sqrt{-g}\bigg[\mathcal{L}_{\rm EH} + \mathcal{L}_\phi + \mathcal{L}_{\phi A} \bigg] \;, \label{eq:action_All}
\end{align}
with
\begin{align}
    \mathcal{L}_{\rm EH} &= \frac{\MP^2}{2} R \;, \\ 
    \mathcal{L}_{\phi A} &=  f(\phi)^2\bigg[- \frac{1}{4} F_{\mu\nu}F^{\mu\nu} - \frac{1}{2}m_A^2 A_\mu A^\mu \bigg]\;, \label{eq:FFphi} \\
    \mathcal{L}_{\phi} &= -\frac{1}{2}(\partial \phi)^2 - V(\phi)\;,
\end{align}
where $R$ is the 4d Ricci scalar, the field strength tensor is defined as $F_{\mu\nu} \equiv \nabla_\mu A_\nu - \nabla_\nu A_\mu$, and $\nabla_\mu$ is the covariant derivative associated with the curved spacetime. The inflaton field and its potential are denoted by $\phi$ and $V(\phi)$, respectively. Throughout this paper, we assume $\phi = \phi(t)$, \textit{i.e.} we are only interested in the dynamics of $A_\mu$ on a fixed background $\phi(t)$. The function $f(\phi)^2$ in Eq.~(\ref{eq:FFphi}) is an arbitrary function of the inflation, which breaks the conformal symmetry of the gauge-field action~\footnote{The breaking of conformal invariance in the gauge sector is typically necessary in order to achieve a sizable amplification of the magnetic field in an expanding universe \cite{Turner:1987bw}.}. Clearly, the presence of the mass term in Eq.~(\ref{eq:FFphi}) breaks the gauge symmetry  under $A_\mu \rightarrow A_\mu + \partial_\mu \Lambda(x)$ transformation, where $\Lambda(x)$ is a scalar function of spacetime coordinates. It is worth stating that the breaking of gauge invariance due to the mass term may be a result of spontaneous symmetry breaking, so that the mass can be regarded as a function of the inflaton. After inflation, we require that the gauge symmetry should be recovered, leading to a vanishing mass term.  

The background metric is assumed to be the spatially flat FLRW metric,
\begin{align}\label{eq:FRW_metric}
    \mathrm{d}s^2 = -\mathrm{d}t^2 + a(t)^2\mathrm{d}\vec{x}^2 \;,
\end{align}
where $t$ is the cosmic time and $a(t)$ is the scale factor. We define the Hubble parameter as $H \equiv \dot{a}/a$ with the dot denoting a derivative with respect to the cosmic time. Note that in this paper we are primarily using the cosmic time. 

Without loss of generality, the vector field $A_\mu$ can be decomposed as $A_\mu = \partial_\mu \chi + A_\mu^{\rm T}$, where $\chi$ represents the longitudinal mode~\footnote{Note that the field $\chi$ in this definition is dimensionless.} and $A_\mu^{\rm T}$ represent the transverse modes. Using such a decomposition in (\ref{eq:FFphi}) it is easy to see that only the $A_\mu^{\rm T}$ modes acquire a mass, while the mode $\chi$ is massless. In order to avoid the $\chi$-mode becoming a ghost, $m_A^2$ has to be positive-definite. Since we are only interested in the dynamics of the gauge field at linear level, the mixing of $\chi$ and $A_\mu^{\rm T}$ modes can be discarded~\footnote{In fact, the mixing term is of the form $A_i^{\rm T} \partial^i \chi$, which, by performing an integration by parts and using $\partial^i A_i^{\rm T} = 0$, does not contribute to the equations of motion.}. Moreover, we will see below that the $\chi$-mode does not contribute to the magnetic field simply because of the fact that $F_{ij}$ only gets a contribution from $A_i^{\rm T}$. For the rest of the paper, we are going to drop the superscript ``${\rm T}$'' and denote the transverse modes as $A_\mu$.

For convenience, we impose the Coulomb gauge condition for the transverse modes: 
\begin{equation}
    A_0(t, \vec{x}) = 0\;, \quad \delta^{ij}\partial_iA_j(t, \vec{x}) = 0\;.
\end{equation}

The Fourier modes, $A_\lambda(t,k)$, of the vector potential are defined via
\begin{align}
    A_i(t, \vec{x}) = \int&\frac{\mathrm{d}^3k}{(2\pi)^3}\sum_{\lambda = \pm}  e^{i\vec{k}\cdot\vec{x}} \textbf{e}^\lambda_i(\hat{k}) \times \nonumber \\
    &\times\left[A_\lambda(t,k) \hat{b}_{\vec{k}}^\lambda + A_\lambda(t,k)^*  \hat{b}_{-\vec{k}}^{\lambda\dagger} \right],
\end{align}
where $k \equiv |\vec{k}|$, $\hat{k} \equiv \vec{k}/k$, and $\textbf{e}^\lambda_i(\hat{k})$ denote the polarization vectors. The creation and annihilation operators are defined to satisfy
\begin{align}
    [\hat{b}^\lambda_{\vec{k}},\hat{b}_{\vec{k}'}^{\lambda \dagger}] = (2\pi)^3 \delta^{\lambda\lambda'} \delta^{(3)}(\vec{k} - \vec{k}') \;.
\end{align}
Note that the polarization vectors satisfy the following orthogonality relations: $\vec{k}\cdot \textbf{e}^\lambda(\hat{k}) = 0$, $\textbf{e}^\lambda(\hat{k}) \cdot \textbf{e}^{\lambda'}(\hat{k}){} = \delta^{\lambda \lambda'}$ and 
$\sum_{\lambda} \textbf{e}_i^\lambda(\hat{k}) \textbf{e}_j^\lambda(\hat{k}) = \delta_{ij} - k_i k_j/k^2$. In addition, due to the symmetries of the FLRW background the mode functions $A_\lambda(t,k)$ depend only on the magnitude $k$ of the comoving momentum.

During inflation, the function $f(\phi)$ can be treated as a function of time. From the action in Eq.~(\ref{eq:action_All}) the equation of motion for the mode function $A_\lambda(t,k)$ is given by
\begin{equation}\label{eq:vector_pot_eom}
    \ddot{A} + \frac{\dot{F}}{F}\dot{A} + \frac{k^2}{a^2} A + m^2_A A = 0 \;,
\end{equation}
where $F \equiv a f^2$, and we have omitted the subscript $\lambda$ since both of the polarizations satisfy the same equation.

In terms of the variable $\mathcal{A} \equiv \sqrt{F} A$ the equation above becomes 
\begin{equation}\label{eq:mode_evolution}
    \ddot{\mathcal{A}} + \left[\frac{k^2}{a^2} + m^2_A + \frac{1}{4}\bigg(\frac{\dot{F}}{F}\bigg)^2 - \frac{\ddot{F}}{2 F}\right]\mathcal{A} = 0 \;.
\end{equation}
This is the main equation we will work with in Section~\ref{sec:toy_model}. It is important to note that in this case the produced EM field is non-helical because the operator $f(\phi)^2 F_{\mu\nu}F^{\mu\nu}$ is even under parity transformations. On the other hand, in the parity-violating scenario considered in e.g. Ref.~\cite{Byrnes:2011aa} the coupling of $\mathcal{A}$ and $\phi$ depends on $k$, leading to suppressed super-horizon enhancement, and causing a back-reaction problem. 

Let us comment on the longitudinal mode $\chi$. As explained before, the mode $\chi$ behaves as a massless scalar field in de-Sitter space and does not contribute to the magnetic field by definition. On top of that, in the energy density we can completely neglect contributions coming from the $\chi$-mode. This can be realized by looking at the equation of motion of $\chi$ in $k$-space:  
\begin{align}
    \frac{d}{dt}(a^3 f^2 \dot{\chi}_{\rm c}) + F k^2 \chi_{\rm c} = 0 \;,
\end{align}
where $\chi_{\rm c} \equiv m_A \chi$. For an oscillating function $F$ considered in Section~\ref{sec:toy_model}, in the super-horizon limit $k \rightarrow 0$ the solutions $\chi_{\rm c}$ would be exponentially decreasing. Therefore, the contributions from the longitudinal mode in the energy density are negligible compared to the ones from the transverse modes.

The electric field $E_\mu$ and magnetic field $B_\mu$ on the FLRW background are defined as
\begin{align}\label{eq:E_B}
    E_{\mu} \equiv F_{\mu\nu} u^\nu \;, \quad B_\mu \equiv \widetilde{F}_{\mu\nu}u^\nu \;,
\end{align}
where $u^\mu$ is the observer's 4-velocity. The dual field strength $\widetilde{F}_{\mu\nu}$ is defined as
\begin{align}
    \widetilde{F}^{\mu\nu} \equiv \frac{1}{2 \sqrt{-g}} \epsilon^{\mu\nu\rho\sigma}F_{\rho \sigma} \;,
\end{align}
where $\epsilon^{\mu\nu\rho\sigma}$ is the totally anti-symmetric Levi-Civita symbol with $\epsilon^{0123}=1$. Using the convention $\epsilon^{0\nu\rho\sigma} = \epsilon^{ijk}$, for a comoving observer with $u^\mu = (1,\vec{0})$ the E- and B-fields defined in Eq.~(\ref{eq:E_B}) become
\begin{align}
    E_\mu = (0, -\dot{A}_i) \;, \quad B_\mu = \bigg(0 , \frac{1}{a}\epsilon_{i}{}^{jk} \partial_j A_k \bigg) \;.
\end{align}
Clearly, from the expression above the longitudinal mode $\chi$ does not give contributions to the B-field~\footnote{Although the field $\chi$ contributes to the E-field, the fact that $\chi$ is just a massless scalar field in an approximately de-Sitter space implies that it does not get amplified outside the horizon.}. The energy density of the gauge field is given by 
\begin{align}
    \rho_{\rm em} &= \langle T^{(A)}_{tt}\rangle = \frac{f^2}{2} \langle E_\mu E^\mu + B_\mu B^\mu + m^2_A A_\mu A^\mu \rangle \nonumber \\
    &\equiv \rho_{\rm E} + \rho_{\rm B} + \rho_{\rm M} \label{eq:rho_tot} \;,
\end{align}
where $T^{(A)}_{\mu\nu}$ is the energy-momentum tensor of the gauge field. The energy densities in the B-field, the E-field and the mass term per logarithmic $k$-interval are given by
\begin{align}
\frac{\mathrm{d}\rho_\mathrm{B}}{\mathrm{d}\log k} &= \frac{1}{4\pi^2}\left(\frac{k}{a}\right)^4 \frac{1}{a} \vert  \sqrt{2k} \mathcal{A}(t, k)\vert^2  \;, \label{eq:rho_B} \\
\frac{\mathrm{d}\rho_\mathrm{E}}{\mathrm{d}\log k} &= \frac{f^2}{4\pi^2}\bigg(\frac{k}{a}\bigg)^2\bigg| \frac{d}{dt}\bigg(\frac{\sqrt{2k}\mathcal{A}(t, k)}{\sqrt{a} f}\bigg)\bigg|^2 \;, \label{eq:rho_E} \\
\frac{\mathrm{d}\rho_\mathrm{M}}{\mathrm{d}\log k} &= \frac{m_A^2}{4\pi^2}\left(\frac{k}{a}\right)^2 \frac{1}{a} \vert  \sqrt{2k} \mathcal{A}(t, k)\vert^2  \;. \label{eq:rho_M} 
\end{align}
In terms of the fractional density parameter $\Omega_{\rm B} \equiv \rho_{\rm B}/\rho_{\rm tot}$, with $\rho_{\rm tot}$ denoting the total energy density of the universe, the present-day  magnetic field strength of $B_0 \approx 10^{-15}~$ Gauss corresponds to $\Omega_{\rm B} \approx 10^{-23}$.

Notice that in the usual case of conformally invariant EM sector Eq.~(\ref{eq:rho_B}) is reduced to
\begin{equation}
    \frac{\mathrm{d}\rho_\mathrm{B}}{\mathrm{d}\log k} = \frac{1}{4\pi^2}\left(\frac{k}{a}\right)^4\cos^2{\bigg(\frac{k}{a H}\bigg)} \approx \frac{1}{4\pi^2}\left(\frac{k}{a}\right)^4 \;. \label{eq:Maxwell_energy}
\end{equation}
where the last term is the leading order in the super-horizon expansion. In this case, the contribution of a particular mode to the energy density rapidly decays as $a^{-4}$ on super-horizon scales. As we will explain in Section~\ref{sec:toy_model}, in our scenario this dilution can be compensated by the exponential growth of the mode functions outside the horizon.

Before concluding this Section, let us comment on the background evolution of the inflaton field. Its dynamics is governed by
\begin{equation}
    \ddot{\phi} + 3H \dot{\phi} + \partial_\phi V = I(E, B, A) \label{eq:phi_EoM}\;,
\end{equation}
where the source term is defined by
\begin{align}
    I(E, B, A) \equiv f \partial_\phi f \langle E_\mu E^\mu - B_\mu B^\mu - m_A^2 A_\mu A^\mu \rangle \;,
\end{align}
representing the back-reaction of the gauge field production on the inflatonary background. This source is related to the difference between the electric and magnetic energy densities and the energy density associated to the mass term, implying that as long as these densities are negligible compared to the inflationary background, the right-hand side of Eq.~(\ref{eq:phi_EoM}) can be safely neglected.

\section{No-go theorem: a review}\label{sec:no-go}

In this Section we briefly review the no-go theorem of the gauge field production during inflation formulated in Ref.~\cite{Byrnes:2011aa}. A crucial assumption of this theorem is that the amplification happens only inside the horizon, while the mode functions freeze when their wavelengths become larger than the Hubble horizon. Since the inflationary energy density stays almost constant, the fractional energy density $\Omega_{\rm B}$ is rapidly diluted as $\propto a^{-4}$ due to the expansion.    

Assuming, for simplicity, that the universe becomes radiation-dominated right at the end of inflation, and stays as such up until present epoch, we can neglect the time evolution of $\Omega_{\rm B}$ after inflation. The observational constraints therefore suggest that the fractional energy density at the end of inflation should satisfy $\Omega^\mathrm{end}_{\rm B} \gtrsim \Omega^\mathrm{obs}_{\rm B} =  10^{-23}$. In passing, it is interesting to note that this lower bound can be relaxed if inflation is followed by an intermediate epoch of stiff fluid domination. Extrapolating $\Omega^\mathrm{end}_{\rm B}$ back when the relevant modes of interest first exited the horizon leads to
\begin{align}\label{eq:Omega_B_nogo}
    \Omega_{\rm B}^\ast = \Omega^\mathrm{end}_{\rm B} e^{4\Delta N_\ast} \;,
\end{align}
where $\Delta N_\ast$ is the number of e-foldings from the horizon exit until the end of inflation. Our primary interest is in the magnetic fields at large scales, corresponding to $\Delta N_\ast \approx 50$, which leads to $\Omega_{\rm B}^\ast \gg 1$; therefore, heavily invalidating the inflationary dynamics. Clearly, this problem is exponentially less severe for small scale production. Indeed, $\Delta N_\ast \approx 11$, corresponding to very small scales, would avoid the back-reaction problem.

For illustration purposes, let us estimate the amplitude of the gauge field $\vert\sqrt{2k}\mathcal{A}\vert$ at the end of inflation required for explaining the large-scale magnetic fields without encountering a large back-reaction during inflation. For this illustration we assume no super-horizon enhancement, and limit ourselves to a more standard situation of conserved mode functions outside the horizon. For simplicity, let us assume that the amplification of $\mathcal{A}$ is peaked at a scale $k_\ast$, $\vert\sqrt{2k}\mathcal{A}\vert^2/a = 2k_\ast\bar{\mathcal{A}}_\ast^2\,\delta(k/k_\ast - 1)$, where $\bar{\mathcal{A}}_\ast$ is a constant amplitude. From Eq.~(\ref{eq:rho_B}), the requirements of no back-reaction ($\rho_B \lesssim \rho_{\phi}$) and satisfying observational limits translate into an inequality 
\begin{align}
    K\left(\Omega^\mathrm{obs}_{\rm B}\right)^{1/2}\lesssim \sqrt{2k_\ast}\bar{\mathcal{A}}_\ast \lesssim e^{-2\Delta N_\ast}K 
     \label{eq:ineq}\;, 
\end{align}
where $K \equiv \sqrt{12}\pi (k_\ast/k_\mathrm{f})^{-2} M_\mathrm{Pl}/H$, with $k_\mathrm{f}$ denoting the horizon scale at the end of inflation. It is clear from this simple estimate that without super-horizon evolution, for typical $\Delta N_\ast \sim 50$ and $\Omega^\mathrm{obs}_{\rm B} = 10^{-23}$, the inequality in Eq.~(\ref{eq:ineq}) cannot be satisfied.

With our mechanism we achieve the mode functions \textit{to be amplified on super-horizon scales,} effectively leading to milder (or complete absence) of dilution. We now proceed to the details of our scenario.

\section{Toy model}\label{sec:toy_model}

\subsection{Analysis}
We demonstrate the main properties of our mechanism in this section. Let us start with discussing the function $F(t)=F(\phi(t))$ introduced in Eq.~(\ref{eq:vector_pot_eom}). We assume that $F(t)$ is an oscillating function with frequency $\omega$ during a certain period of time. It is then convenient to introduce the variable $z \equiv \omega(t - t_{\rm i})/2$, where $t_i$ denotes the onset of oscillations. For simplicity, we assume an exponentially expanding universe with $a(z) = \exp(2 r z)$, where $r \equiv H/\omega$.

A simple model to ensure the positive-definiteness of $F$ is given by 
\begin{align}\label{eq:Function_F}
     \frac{F^\prime}{F} = -2 \gamma \sin(2z)\;.
 \end{align}
This choice renders the mode-function equation Eq.~(\ref{eq:mode_evolution}) to take the form of Whittaker-Hill equation \cite{whittaker_watson_1996} with a time-dependent leading coefficient,
\begin{align}\label{eq:EoM_cosmic_time}
    \mathcal{A}^{\prime\prime} + \bigg[C(z) + 2 q \cos(4z) + 2 p \cos(2 z) \bigg]\mathcal{A} = 0 \;,
\end{align}
where the prime denotes a derivative with respect to $z$. Similar equation has been analyzed in Ref.~\cite{Enqvist:2016mqj} in the context of vacuum stability in Higgs-inflation, and in Ref.~\cite{Lachapelle:2008sy} in the context of preheating with non-standard kinetic terms. We provide a brief review of such equations in Appendix~\ref{app:Hill_equation}. In Eq.~(\ref{eq:EoM_cosmic_time}) all the coefficients are fixed in terms of a single parameter $\gamma$,
\begin{align}\label{eq:para_toy}
    C(z) \equiv \frac{4k^2}{a^2 \omega^2} - \frac{\gamma^2}{2} + \delta\;, \quad p \equiv \gamma \;, \quad q \equiv \frac{1}{4}\gamma^2 \;,
\end{align}
where $\delta \equiv 4m_A^2 r^2/H^2$.

It is useful to note that in the case where $m_A^2 = 0$ the mode function $A$ satisfies $(F \dot{A})^{\boldsymbol{\cdot}}  = 0$ in the limit $k/a \rightarrow 0$. This implies that when $F$ is assumed to be a bounded function, the massless vector field cannot be exponentially amplified on super-horizon scales, see e.g. \cite{Shtanov:2020gjp} for a similar discussion. This means that the mass term plays an important role in our analysis. For exponentially decaying $F$ the mode functions $A$ can be amplified on super-horizon scales. We will comment on this possibility and its drawback in Section~\ref{subsec:connec_obs}.

In order to derive  analytical approximations, we will neglect for now the time-dependent term in $C(z)$ (achieved in the super-horizon limit), and will rely on the Floquet theorem \cite{whittaker_watson_1996,McLachlanN.W.NormanWilliam1964Taao}. The general solution of Eq.~(\ref{eq:EoM_cosmic_time}) can be written as 
\begin{align}\label{eq:mode_analytic}
    |\sqrt{2 k}~\mathcal{A}(z)| = e^{\pm \mu z} h(z) \;,
\end{align}
where $\mu$ is the Floquet exponent and $h(z)$ is a periodic function with period $\pi$. The growth rate $\mu$ can be analytically determined using Eq.~(\ref{eq:mu_R}), and it depends on the parameters $\gamma$ and $\delta$; see also the discussion in Appendix~\ref{app:Hill_equation}. Fig.~\ref{fig:mu_gamma_main} demonstrates the dependence of $\mu$ on the parameter $\gamma$, where we fix $\delta = 10^{-2}$. We use this result for guiding our selection of valid parameter combinations. Particularly, $\gamma = 7$ corresponds to large super-horizon enhancement with growth rate of $\mu \approx 2.5$.  

\begin{figure}[t!]
    \centering
    \includegraphics[width=\columnwidth]{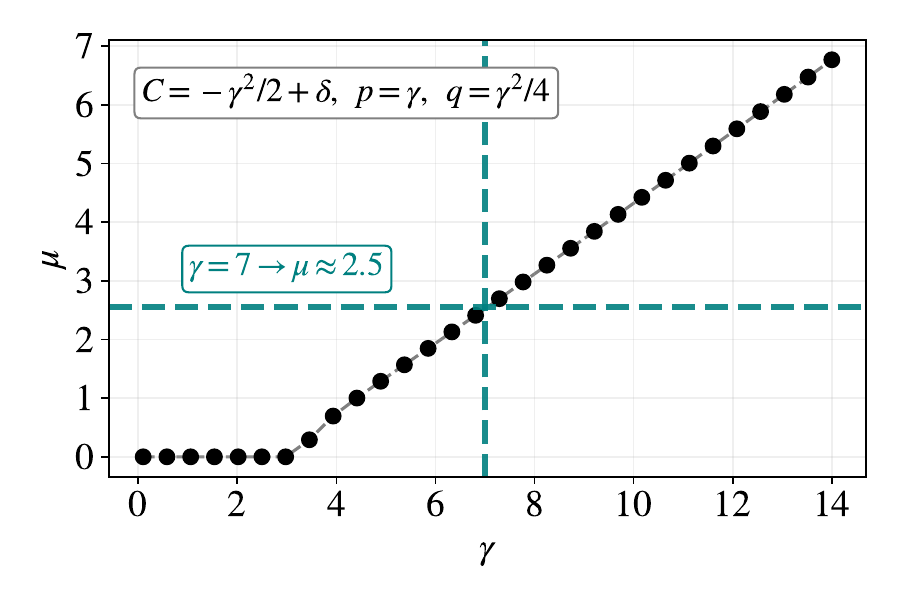}
    \caption{Dependence of the growth exponent $\mu$ on the parameter $\gamma$, for $\delta = 10^{-2}$. The intersection of the two dashed lines marks the parameter choice leading to scale-invariant magnetic spectrum. The $\mu(\gamma)$ scaling is well-described by the simple fitting function provided in Appendix~\ref{app:Hill_equation}, and is approximately given by $\mu = max\left\{0, 0.6\gamma - 1.64\right\}$.}
    \label{fig:mu_gamma_main}
\end{figure}

\begin{figure}[ht!]
    \centering
    \includegraphics[width=\columnwidth]{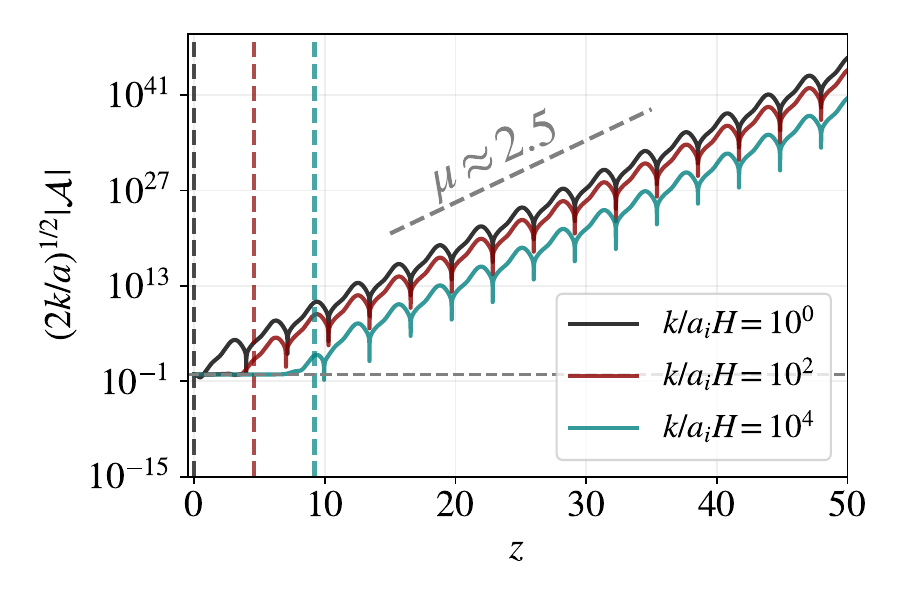}
    \caption{Numerical solutions of $\mathcal{A}$ for different scales $k$. The vertical lines mark the horizon crossing of the corresponding mode. Here, $\gamma = 7$, $r = 0.5$, $\delta = 10^{-2}$ and $\Delta N = N_{\rm f} - N_{\rm i} = 50$.}
    \label{fig:Cosmic_time_Sol}
\end{figure}

We now turn to exact numerical analysis of Eq.~(\ref{eq:EoM_cosmic_time}). We particularly solve it in the range spanning from $z_{\rm i} = 0$ to $z_{\rm f} = 50$, which, in terms of e-folding number $\Delta N$ is given by $z_{\rm f} - z_{\rm i} = \Delta N/(2r)$. We impose the following initial conditions:
\begin{align}\label{eq:initial_con}
    \vert\sqrt{2 k}~\mathcal{A}(z)\vert_{z = z_{\rm i}} = 1 \;, \quad \bigg|\sqrt{2 k}~\frac{\mathrm{d}\mathcal{A}(z)}{\mathrm{d}z}\bigg|_{z = z_{\rm i}} = \frac{2k}{\omega} \;.
\end{align}
The numerical solutions are shown in Fig.~\ref{fig:Cosmic_time_Sol}. Notice that the mode with momentum $k$ crosses the horizon at 
\begin{align}
    z_c = \frac{1}{2r}\log\bigg(\frac{k}{\omega r}\bigg) \;,
\end{align}
which is marked by vertical dashed lines in Fig.~\ref{fig:Cosmic_time_Sol}. It is evident from Fig.~\ref{fig:Cosmic_time_Sol} that the mode functions experience a large amplification once they cross the horizon, leading to gauge field production at super-horizon scales. Note that here we have chosen $\gamma = 7$ and $\delta = 10^{-2}$, and the resulting growth agrees with the prediction of Fig.~\ref{fig:mu_gamma_main}.  

It is interesting to note that a similar amplification could have been achieved in a model with tachyonic mass. In our scenario, however, we avoid the usual ghost and gradient pathologies encountered in tachyonic gauge theories since the coefficients of kinetic terms do not change signs during the evolution. 

\subsection{Phenomenological validity}\label{subsec:connec_obs}

Having computed the mode functions we can now estimate the magnetic field spectrum. First, notice that the mode functions can be approximated as $\vert\sqrt{2k}\mathcal{A}\vert \sim \exp(\mu \Delta z_k)$, where the growth factor $\mu$ is approximately universal for all the modes; see Fig.~\ref{fig:Cosmic_time_Sol}. The duration of amplification $\Delta z_k$, on the other hand, clearly depends on the mode since the enhancement is effectively active only when the given mode becomes super-horizon. We can therefore write
\begin{align}
    \Delta z_k = \frac{1}{2r}\log \bigg(\frac{k_{\rm f}}{k}\bigg) \;,
\end{align}
where  $k_{\rm f}$ is the comoving horizon scale at the end of inflation. Clearly, for $k = k_{\rm f}$ we have $\Delta z_{k} = 0$ and there is no amplification for this mode. 

From Eq.~(\ref{eq:rho_B}) we have
\begin{align}\label{eq:OmegaB_obs}
 \frac{\mathrm{d}\Omega_{\rm B}}{\mathrm{d} \log k}  = \mathcal{C} \cdot 10^{-12} \bigg(\frac{H}{10^{13}~{\rm GeV}}\bigg)^2 \bigg(\frac{k}{k_{\rm f}} \bigg)^{5 - \mu/r}\;,
\end{align}
where $\mathcal{C}$ is an $\mathcal{O}(1)$ numerical factor. The fifth power of $k$ emerges due to the $a^{-5}$ factor.

It is now easy to infer the spectral tilt from Eq.~(\ref{eq:OmegaB_obs}). Particularly, for $5 - \mu/r < 0$ , the spectrum is red-tilted, \textit{i.e.} it gets large contributions from small-$k$ modes. This is because the exponent $\mu$ of the mode functions overcompensates the dilution factor ($a^{-5}$) due to the expansion of the universe. On the other hand, for $5 - \mu/r > 0$, the spectrum is blue-tilted because in this case the super-horizon dilution dominates over the super-horizon growth of mode functions. The exact balance between dilution and amplification is achieved when $5 - \mu/r = 0$, which therefore leads to a scale-invariant energy spectrum.

\begin{figure}[t]
    \centering
    \includegraphics[width=\columnwidth]{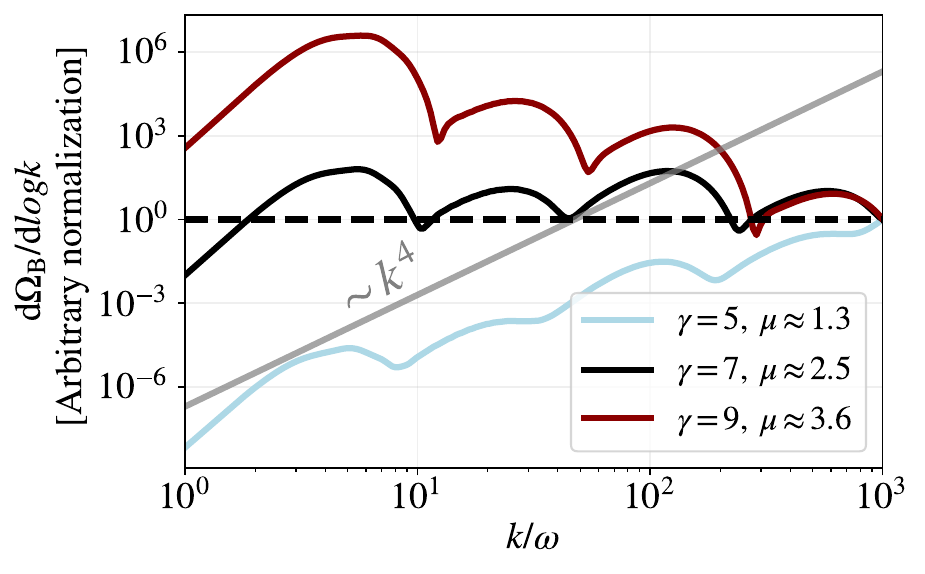}
    \caption{Numerical evaluation of the magnetic energy spectra at the end of inflation. The spectra are arbitrarily normalized in order to facilitate the comparison of their shapes. The spectrum is red-tilted for $\mu \approx 3.6$, blue-tilted for $\mu \approx 1.3$, and scale-invariant for $\mu \approx 2.5$ -- all in agreement with our analytical result of Eq.~(\ref{eq:dOmega_B_obs}). The gray solid line represents the spectrum in the Maxwell limit, and the black dashed line represents the scale-invariant spectrum for guiding the eye. We used $r = 0.5$, $\delta = 10^{-2}$ and $\Delta N = 50$.}
    \label{fig:dOmega_B}
\end{figure}

The usual energy spectrum in the $U(1)$-symmetric EM theory is recovered when $\mu/r = 1$ in Eq.~(\ref{eq:OmegaB_obs}). This is a consequence of our variable choice in the cosmic-time frame, \textit{i.e.} $\mathcal{A} = \sqrt{F} A$, which grows as $\sqrt{a}$ outside the horizon. In this case the spectrum is very blue-tilted and scales as $k^4$; see Eq.~(\ref{eq:Maxwell_energy}).

In addition to our analytical result in Eq.~(\ref{eq:OmegaB_obs}), in Fig.~\ref{fig:dOmega_B} we show the magnetic field spectra calculated using the exact numerically-evaluated mode functions. The gray solid line represents the spectrum in the Maxwell limit, while the red, blue and black solid lines represent red-tilted, blue-tilted and scale-invariant spectra, all agreeing with our analytical expectations. The spectra are arbitrarily normalized for a more convenient comparison of their slopes.   

\begin{figure}[ht!]
    \centering
    \includegraphics[width=\columnwidth]{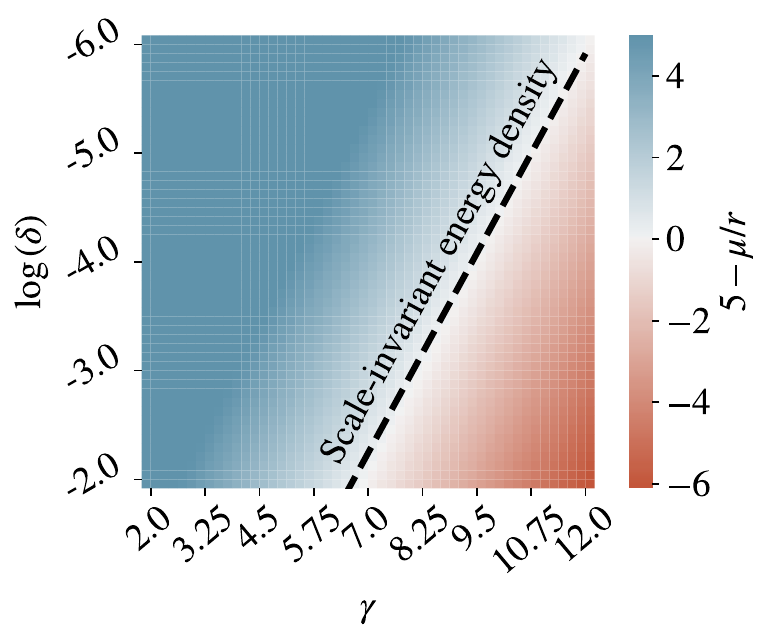}
    \caption{The power index of the spectrum $5 - \mu/r$ in the $\delta$--$\gamma$ parameter space. The dashed line represents the scale-invariant magnetic energy density: $5 - \mu/r = 0$.}
    \label{fig:param_space}
\end{figure}

In order to explore a wider parameter range, in Fig.~\ref{fig:param_space} we demonstrate the dependence of $\mu/r$ on $\gamma$ and $\delta$. The diagonal dashed line represents the parameter combinations resulting in a scale-invariant magnetic spectrum.

It is useful to rewrite Eq.~(\ref{eq:OmegaB_obs}) by introducing a typical CMB pivot scale of $k_{\rm CMB} \sim 0.01~{\rm Mpc^{-1}}$. We obtain  
\begin{align}
  \frac{\mathrm{d}\Omega_{\rm B}}{\mathrm{d} \log k} \simeq 10^{-12 + \xi} \bigg(\frac{H}{10^{13}~{\rm GeV}}\bigg)^2 \bigg(\frac{k}{{\rm Mpc^{-1}}} \bigg)^{5 - \mu/r}  \;, \label{eq:dOmega_B_obs}
\end{align}
where we have omitted the $\mathcal{O}(1)$ numerical factor and have introduced $\xi \equiv (5r - \mu)(2 - \Delta N_{\rm CMB})/(r \log 10)$, with $\Delta N_{\rm CMB}$ being the inflationary e-folding number corresponding to the typical CMB scales, $\Delta N_{\rm CMB} \equiv \log(a_{\rm f}/a_{\rm CMB})$. As an example, fixing $H \sim 10^{13}~{\rm GeV}$ and $\Delta N_{\rm CMB} = 50$, the requirement $\Omega_{\rm B}(z_{\rm f}) \gtrsim 10^{-23}$ at the scale of $k \approx \mathcal{O}(1){\rm Mpc^{-1}}$ results in a lower bound
$\mu/r \gtrsim 4.47$. We will show, however, that a working scenario requires a lower energy scale. More generally, the requirement that $\Omega_{\rm B} \gtrsim 10^{-23}$ at the end of inflation can be reformulated as follows. From Eq.~(\ref{eq:OmegaB_obs}), choosing $k = k_{\rm B}$ (the scale relevant to the magnetic-field production) we obtain the following inequality:
\begin{align}
    2x + n_{\rm B} (\kappa - \Delta N_{\rm CMB}) \gtrsim -23\log(10) \;. \label{con:OmegaB1}
\end{align}
where $n_{\rm B} \equiv 5 - \mu/r$, $\kappa \equiv \log(k_{\rm B}/k_{\rm CMB})$, $x \equiv \log(H/\MP)$ and $x_0 \equiv \log(H_0/\MP)$, with $H_0$ being the Hubble parameter today.

Next we will discuss three important aspects of our proposal.

\paragraph{\textbf{No $\Omega_{\rm B}$ and $\Omega_{\rm M}$ back-reaction.}}

In addition to the observational bound in Eq.~(\ref{con:OmegaB1}), we should also make sure that gauge field enhancement does not lead to significant back-reacktion, \textit{i.e.} $\Omega_{\rm B} < 1$ and $\Omega_{\rm M} < 1$. The first of these gives us:
\begin{align}
    2x + n_{\rm B} (\kappa - \Delta N_{\rm CMB}) \lesssim 0 \;.  \label{con:OmegaB2}
\end{align}

Let us now consider the back-reaction from $\Omega_{\rm M}$. From Eq.~(\ref{eq:rho_M}) and the mode function solutions we obtain
\begin{align}\label{eq:Omega_M}
    \frac{d\Omega_{\rm M}}{d\log k} \simeq \frac{\delta}{r^2} \bigg(\frac{H}{\MP}\bigg)^2 \bigg(\frac{k}{k_{\rm f}}\bigg)^{n_\mathrm{B} - 2} \;,
\end{align}
where we have omitted an unimportant $\mathcal{O}(1)$ numerical prefactor. Even in the case of a scale-invariant magnetic spectrum the above energy contribution is red-tilted, suggesting a potential back-reaction problem. As we mentioned earlier, the mass term $\delta$ is essential for our resonant mechanism to operate, and we cannot exponentially lower this term. In fact, recall that for the scale-invariant magnetic spectrum we need $\delta/r^2 = \mathcal{O}(1)$ when $\gamma$ is chosen to be moderately small. Alternatively, we could lower the energy-scale of inflation, as well as lower the inflationary e-folding number corresponding to the scales of interest for magnetic field generation. The inflationary scale, however, cannot be lowered arbitrarily.

From Eq.~(\ref{eq:Omega_M}) we can derive the condition to avoid the back-reaction from $\Omega_{\rm M}$, $\Omega_{\rm M} < 1$, as 
\begin{align}
     2x + (n_{\rm B} - 2) (\kappa - \Delta N_{\rm CMB}) \lesssim 0 \;. \label{con:OmegaM}
\end{align}
Note that for this estimate we have fixed $\delta/r^2$ to be of order unity.

Finally, $\Delta N_{\rm CMB}$ is fixed in terms of $H/\MP$ by requiring a long-enough inflationary period, $(a_0 H_0)^{-1} < (a_{\rm CMB} H)^{-1}$. Assuming an instant transition to radiation-dominated universe we have 
\begin{align}\label{eq:suc_inflation}
    \Delta N_{\rm CMB} \gtrsim \frac{1}{2} (x - x_0) \;.
\end{align}

For simplicity, we use $\Delta N_{\rm CMB} = \frac{1}{2} (x - x_0)$, so that the constraints in Eqs~(\ref{con:OmegaB1}), (\ref{con:OmegaB2}) and (\ref{con:OmegaM}) become
\begin{align}
    (4 - n_{\rm B})x + n_{\rm B} (2 \kappa + x_0) &\gtrsim -46\log(10) \;, \label{eq:con:OmegaB1} \\ 
    (4 - n_{\rm B})x &\lesssim -n_{\rm B} (2 \kappa + x_0) \;, \label{eq:con:OmegaB2} \\ 
    (6 - n_{\rm B})x &\lesssim -(n_{\rm B} - 2) (2 \kappa + x_0) \;. \label{eq:con:OmegaM} 
\end{align}
For a fixed value of $\kappa$, the inflationary scale $H/\MP$ and the magnetic spectral tilt $n_{\rm B}$ should be chosen carefully in order to satisfy all the above constraints.

\begin{figure}[ht!]
    \centering
    \includegraphics[width=\columnwidth]{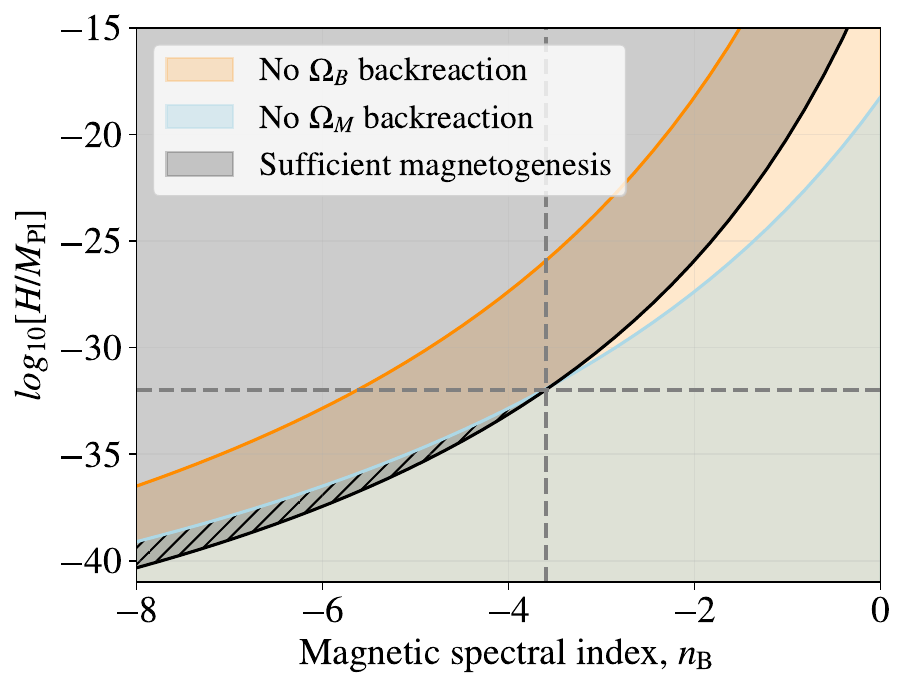}
    \caption{Parameter regions where our mechanism produces sufficiently large magnetic fields, and avoids strong backreaction from the magnetic and potential energy sectors. The intersection, denoted by slanted gray shading, corresponds to the region satisfying Eqs.~(\ref{eq:con:OmegaB1}), (\ref{eq:con:OmegaB2}) and (\ref{eq:con:OmegaM}).}
    \label{fig:hubble_nB}
\end{figure}

In Fig.~\ref{fig:hubble_nB}, we plot these three constraints and show that the allowed region is in the range $n_{\rm B} \lesssim -4$ and $H/\MP \lesssim 10^{-32}$ where we have fixed $\kappa = 7$~\footnote{Note that $\kappa = 7$ exactly corresponds to \textit{the scales of our interest for large-scale magnetic field generation, \textit{i.e.} $k_{\rm B} = 10^3k_{\rm CMB}$.}} and $x_0 = -140$. The reheating temperature associated to the allowed inflationary scale $H/\MP \sim 10^{-32}$ is approximately $T_{\rm R} \sim 10^3~{\rm GeV}$. It should be noted that $n_{\rm B} \approx -4$ is supported by our model, as can be seen from Fig.~\ref{fig:param_space}. In addition, we clearly see from Fig.~\ref{fig:hubble_nB} that the scale-invariant case with $n_{\rm B} = 0$ is outside the allowed region.

In Fig.~\ref{fig:method_comparison} we plot the exponent $\mu$ versus $\gamma$ for different values of $\delta$, which indicates that the value of $\delta$ can be lowered by choosing a much larger $\gamma$, potentially widening the available parameter space. We thus conclude that in order for our resonant magnetogenesis to explain the red-tilted $\Omega_{\rm B}$ without an excessive back-reaction from $\Omega_{\rm M}$ we need to consider a low-scale inflation.

For completeness let us also estimate the effect of an intermediate transition epoch in between inflation and radiation-domination. Depending on the equation of state during such a transition epoch $\Omega_\mathrm{B}$ could be further amplified if the universe is dominated by a stiff fluid. Such scenarios could naturally happen in models connecting inflation and dark energy; see e.g. Refs.~\cite{Akrami:2017cir, Rubio:2017gty} for well-motivated examples.   

For the sake of a more general argument let us assume that in the transition epoch lasting for $\Delta N_{\rm tr}$ e-foldings the Hubble function evolves as $H \sim a^{-\beta}$ with $\beta$ being a constant. During this epoch we have $\Omega_{\rm B} \sim a^{2(\beta - 2)}$, meaning that depending on the value of $\beta$, $\Omega_{\rm B}$ can either get further enhanced or be suppressed. We assume that the mass term vanishes right after inflation, recovering the standard gauge-invariant Maxwell theory in the late universe, and there is no production of $\Omega_{\rm M}$ during the transition phase. The conditions in Eqs.~(\ref{eq:con:OmegaB1})--(\ref{eq:con:OmegaM}) are modified to
\begin{align}
    (4 - n_{\rm B})x + n_{\rm B} (2 \kappa + x_0) + 46\log(10) + Q  &\gtrsim 0  \;, \label{eq:con:OmegaB1_Tr} \\ 
    (4 - n_{\rm B})x  + n_{\rm B} (2 \kappa + x_0) + Q &\lesssim 0 \;, \label{eq:con:OmegaB2_Tr} \\ 
    (6 - n_{\rm B})x + (n_{\rm B} - 2) \left(2 \kappa + x_0 + \frac{Q}{n_{\rm B} - 4}\right) &\lesssim  0\;. \label{eq:con:OmegaM_Tr} 
\end{align}
where $Q \equiv \Delta N_{\rm tr} (\beta - 2)(4 - n_{\rm B})$. Clearly, by setting $\Delta N_{\rm tr} = 0$ we can recover Eqs.~(\ref{eq:con:OmegaB1})--(\ref{eq:con:OmegaM}). We have checked that in case of $\beta = 3$ (kination period) in a transition epoch lasting several e-foldings the gray wedge in Fig.~\ref{fig:hubble_nB} can be shifted toward less-negatively-tilted spectra and to higher Hubble-scales during inflation.

\paragraph{\textbf{No strong coupling.}}

The inverse of the function $f(\phi)$ determines the effective gauge coupling in our model and it should always remain smaller than unity in order for our perturbative treatment to be valid. The normalization of $f$ itself, however, does not appear in our model, allowing us to rescale $f$ to arbitrary values. 

It is useful to note that in the case of standard parametrization where $f \sim a^\alpha$, the scale-invariant magnetic spectrum is achieved with $\alpha = 2$ \cite{Bamba:2006ga}, corresponding to a rapidly growing $f$. Requiring $f^{-1}$ to be smaller than unity during approximately the last $50$ e-foldings of inflationary epoch leaves us with $f \gtrsim e^{100}$ at the end of inflation. This corresponds to extremely weakly-coupled plasma in the post-inflationary universe. A separate mechanism is therefore required to effectively lower the value of $f$ from $e^{100}$ to $\mathcal{O}(1)$, posing an interesting challenge.

In our scenario $f$ evolves mildly; for an approximately constant $F$~\footnote{From Eq.~(\ref{eq:Function_F}) it is easy to realize that $F(z)$ is an oscillating function regardless the values of $\gamma$. One can always choose the normalization of $F(z)$ such that it is oscillating around unity during inflation.}, we have $f \sim a^{-1/2}$. Not only is this a milder evolution compared to the $f \sim a^2$ case, but it is also a decreasing function. This is a noteworthy property, because in our scenario we can \textit{start} with a very weakly-coupled theory, and dynamically evolve toward $f \sim \mathcal{O}(1)$ at the end of inflation.

It is interesting to note in this context that the form of the source term in Eq.~(\ref{eq:EoM_cosmic_time}) required for a successful resonance can be realized without an explicit mass term for the gauge field. Such a possibility can be realized by requiring the function $\tilde{f} = \sqrt{F}$ to satisfy the following differential equation:
\begin{align}
    \tilde{f}'' + \bigg[-\frac{\gamma^2}{2} + \delta + 2 q \cos(4z) + 2 p \cos(2z) \bigg] \tilde{f} = 0 \;. \label{eq:EoM_tilde_f}
\end{align}
This equation by itself is of the Whittaker-Hill form, and not only does this lead to an exponentially evolving amplitude of the coupling function, but also leads to oscillations in the latter. When the coupling function crosses zero during oscillations, the gauge coupling becomes infinite, invalidating such a possibility.

\paragraph{\textbf{Screening of the electric fields.}}
The electric sector can also be amplified due to the exponential growth of the gauge-field mode functions during inflation. Interestingly, both $\Omega_{\rm M}$ and $\Omega_{\rm E}$ scale similarly (see Eqs.~(\ref{eq:rho_E})--(\ref{eq:rho_M})), and our results for keeping $\Omega_{\rm M}$ back-reaction under control would automatically imply no strong electric field back-reaction. 

An additional argument regarding the suppression of electric fields during inflation is related to the Schwinger pair production. When exceeding a certain critical value, the electric field produces electron-positron pairs \cite{Schwinger:1951nm,Dunne:1998ni}, whose field can screen the original electric field (see also Refs.~\cite{Chu:2010xc,Gold:2020qzr} for an attempt to describe the dynamics of the process in a flat space). Limiting ourselves to a flat space for simplicity, the pair creation rate in an external electric field $E$ scales as $\exp[-\pi m_\mathrm{e}^2/(e E)]$, where $m_\mathrm{e}$ is the electron mass and $e$ is its charge. Qualitatively, above a critical field value of $E_\mathrm{crit} \gtrsim 2m_{\rm e}^2/e$, the energy of the electric field in a region of size $\sim 1/m_\mathrm{e}$ exceeds the rest-mass energy of an electron-positron pair, leading to effective pair production~\footnote{In order to make fermions massive during inflation one could consider a non-minimal coupling of the fermions and the curvature, generating an effective electron mass $m_{\rm e}^2 \gg H^2$.}. The produced electron-positron pairs then induce an electric field which screens the effect of the external (amplified) electric field, ameliorating the back-reaction risk. In our scenario, the electric field is amplified way beyond the required critical value, therefore leading to the screening described above. For this qualitative picture we assume that the pair-induced electric field is homogeneous and static on large scales and does not affect the amplification of the magnetic field. A more detailed analysis might reveal that the produced pairs could also generate currents and affect the magnetic fields. We leave such an analysis for future.

\section{Conclusions}\label{sec:conclusion}
In this work, we have proposed a novel inflationary  mechanism for generating the observed large-scale cosmological magnetic fields in the late universe. In contrast to the existing similar proposals, which considered sub-horizon amplification, our mechanism relies on the super-horizon enhancement of the gauge-field mode functions. This crucial difference allows our model to evade the back-reaction no-go theorem of Ref.~\cite{Byrnes:2011aa}. In our scenario, the evolution of the mode functions is determined by the Whittaker-Hill equation, leading to the parametric resonance for the modes when they exit the comoving Hubble horizon. We have presented a detailed analysis of the resonance, obtaining the numerical and analytical solutions to the Whittaker-Hill equation in the regime not previously discussed in the literature. Particularly, the semi-analytical method explained in Appendix~\ref{app:Hill_equation} can be applied to other physical systems which are governed by similar equations. 

Further, we have evaluated the spectrum of the magnetic field energy density $\Omega_{\rm B}$, finding that our model is able to naturally explain scale-invariant, as well as red- and blue-tilted spectra. We have demonstrated that the observational limit of $\Omega_{\rm B}^{\rm obs} \gtrsim 10^{-23}$ can be easily satisfied, without the danger of back-reactions on the inflationary background from the enhanced magnetic sector. We have also argued that super-horizon resonance amplification requires a mass term for the gauge field, and have shown that the back-reaction for this additional potential-energy sector can be neglected if the magnetic spectrum is sufficiently red-tilted. Similarly, we have presented arguments suggesting that there is no strong back-reaction from the electric sector.        

Additionally, we have argued that in contrast to many conformal-invariance-breaking scenarios presented in the literature, our model does not suffer from the drastic evolution of the coupling $f$. The function $F$ is an oscillating function and does not need to increase exponentially in order to guarantee a sufficiently sizeable enhancement of the magnetic field.

This work can be extended in several interesting directions. Particularly, we leave a detailed comparison with observational constraints to a future work. Second, it would be interesting to consider the same mechanism in related contexts. For instance, during preheating the inflaton background is usually assumed to be an oscillating function of time, which might result in further enhancement of the magnetic energy density. Finally, it is worth investigating relations between magnetic fields and the produced gravitational waves in related scenarios; see for example Ref.~\cite{Cai:2021yvq} which studies the resonant production of gravitational waves during inflation.

\section*{Acknowledgements}
We thank M.~Ata, T.~Fujita, A.~Kusenko, C.~Lin and Y.~Shtanov for useful discussions. M.S., V.V. and V.Y. are supported by the WPI Research Center Initiative, MEXT, Japan. The work of M.S. has been additionally supported by JSPS KAKENHI grants 19H01895,  20H04727,  20H05853, V.V. has been supported by JSPS KAKENHI grant 20K22348, V.Y. has been supported by JSPS KAKENHI grant JP22K20367.

\appendix

\section{Whittaker-Hill equation}\label{app:Hill_equation}

\begin{figure*}[ht!]
    \centering
    \includegraphics[width=\textwidth]{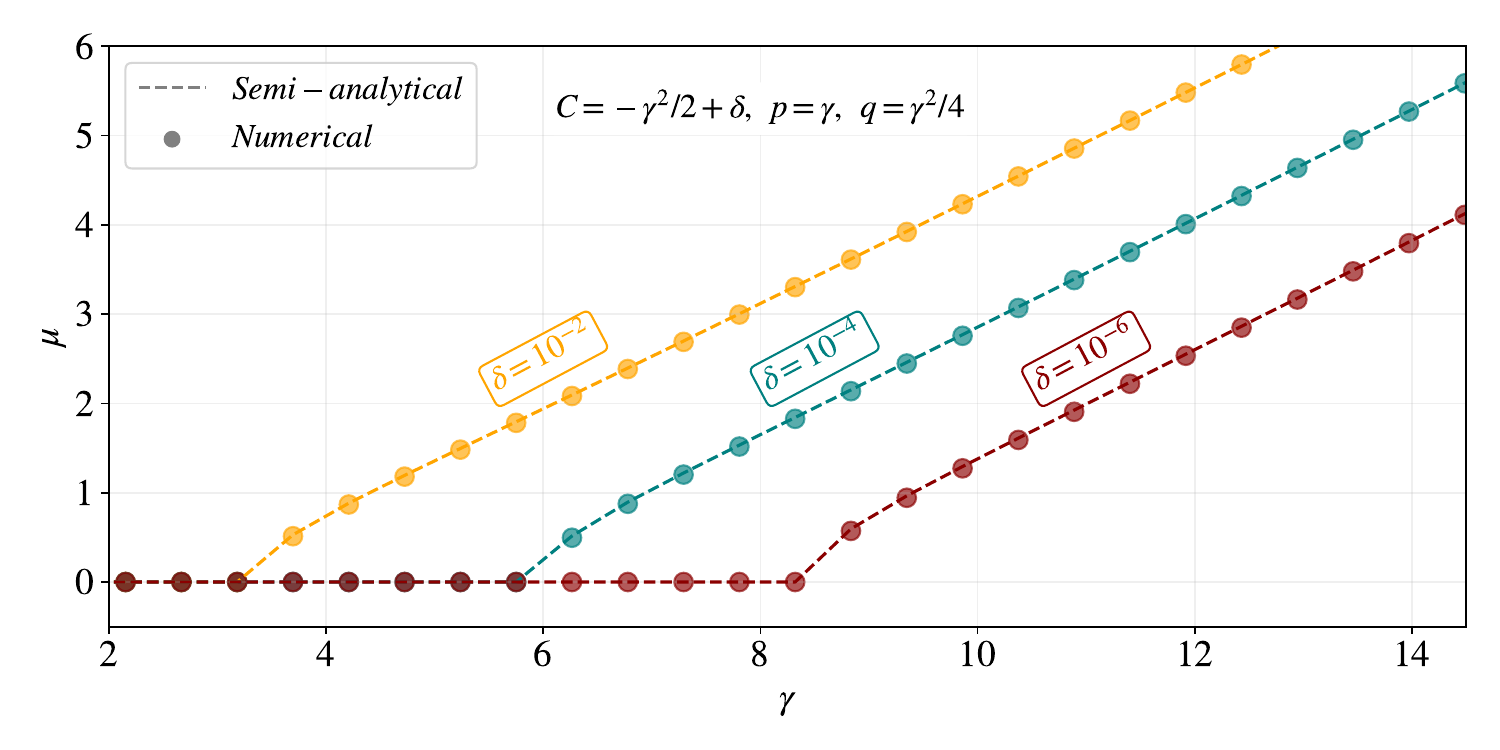}
    \caption{The growth rate $\mu$ as a function of $\gamma$ for different values of the offset parameter $\delta$. In the parameter range $\delta \in [10^{-6}, 10^{-2}]$, a very good approximation to these solutions is given by a simple fitting function $\mu = max\left\{0, 0.6\gamma - 0.85\vert\log_{10}\delta\vert^{0.95}\right\}$.}
    \label{fig:method_comparison}
\end{figure*}

In this Appendix we provide a brief review of the Whittaker-Hill equation (see Ref.~\cite{whittaker_watson_1996} for an in-depth exposition) and will provide the relevant expressions regarding the (in)stability of its solutions. The Whittaker-Hill equation is of the form,
\begin{align}\label{eq:A_eom_two_F}
    \frac{d^2\mathcal{A}}{dz^2} + \bigg[C + 2 q \cos(4z) + 2 p \cos(2 z) \bigg]\mathcal{A} = 0 \;.
\end{align}

According to the Floquet theorem the solution of (\ref{eq:A_eom_two_F}) can be written as
\begin{align}
    \mathcal{A}(z) = e^{\mu z} \sum_{n = -\infty}^\infty c_{2n} e^{2inz} \;,
\end{align}
with constant $c_{2n}$ coefficients. The Floquet exponent $\mu$ can be computed using the following expression \cite{whittaker_watson_1996}:
\begin{align}\label{eq:mu_Hill}
    \mu = -\frac{i}{\pi} \arccos\bigg[1 + \Delta(0)\bigg(\cos(\pi\sqrt{C}) - 1\bigg)\bigg] \;,
\end{align}
where $\Delta(i \mu)$ is the determinant of an infinite matrix:
\begin{align}\label{eq:infinite_matrix}
    \Delta(i\mu) \equiv \left\lvert
\begin{matrix}
\ddots &  &  & & & & & & \\
 & \tilde{\zeta}_{-2} & \zeta_{-2} & 1 & \zeta_{-2} & \tilde{\zeta}_{-2} & 0 & 0 & \\
 & 0 & \tilde{\zeta}_{0} & \zeta_{0} & 1 & \zeta_{0} & \tilde{\zeta}_{0} & 0 & \\
  & 0 & 0 & \tilde{\zeta}_{2} & \zeta_{2} & 1 & \zeta_{2} & \tilde{\zeta}_{2} & \\
 &  &  & & & & & & \ddots
\end{matrix}
\right\rvert \;,
\end{align}
and 
\begin{align}
  \zeta_{2n} \equiv \frac{p}{C - (i\mu - 2n)^2} \;, \quad \tilde{\zeta}_{2n} \equiv \frac{q}{C - (i\mu - 2n)^2} \;.
\end{align}
In general, the exponent $\mu$ is a complex number, $\mu = \mu_R + i\theta$ with $\mu_R, \theta \in \mathbb{R}$. The solution is unstable if $\mu$ has a non-zero real part ($\mu_R \neq 0$).

For negative $C$, using Eq.~(\ref{eq:mu_Hill}), we obtain
\begin{align}\label{eq:find_mu_R}
    \alpha \cosh(\pi \mu_R) = 2 \Delta(0) \sinh^2(\frac{\pi}{2}\sqrt{|C|}) + 1 \;,
\end{align}
where $\alpha = \pm1$ depending on the sign of the right-hand side. Since $\Delta(0)$ is a real number for any real $\gamma$, $\theta$ should be an integer. Thus, using Eq.~(\ref{eq:find_mu_R}), the solution is given by
\begin{align}\label{eq:mu_R}
    \mu_R = \frac{1}{\pi}\log\bigg[\alpha D \pm \sqrt{D^2 - 1}\bigg] \;,
\end{align}
where $D \equiv 2 \Delta(0) \sinh^2(\frac{\pi}{2}\sqrt{|C|}) + 1$. Notice that there are two roots for each sign of $\alpha$ resulting in the same value of $\vert\mu_R\vert$. Later on, we use $\mu$ to denote the real part $\mu_R$.

In order to find the growth rate $\mu$, the determinant of an infinite matrix in Eq.~(\ref{eq:infinite_matrix}) should be evaluated. However, in practice, the matrix can be truncated at a finite size, even though large truncation size could be necessary around the boundaries of instability regions, rendering the evaluation of the determinant with standard methods technically challenging. The special structure of the matrix allows it to be considered as a tri-block-diagonal matrix, for which the determinant can be computed recursively \cite{SALKUYEH2006442}. Fig.~\ref{fig:method_comparison} shows the values of $\mu$ as a function $\gamma$ for different values of $\delta$, and we see that the numerical values of $\mu$ (denoted by dots) agree with the semi-analytical results (dashed lines). For completeness, we give an simple fitting function for $\mu$ as a function of $\gamma$ and $\delta$: $\mu(\gamma,\delta) = max\left\{0, 0.6\gamma - 0.85\vert\log_{10}\delta\vert^{0.95}\right\}$.


\bibliographystyle{utphys}
\bibliography{bibliography}

\providecommand{\noopsort}[1]{}\providecommand{\singleletter}[1]{#1}%
\providecommand{\href}[2]{#2}\begingroup\raggedright\begin{thebibliography}{10}

\bibitem{Ando:2010rb}
S.~Ando and A.~Kusenko, ``{Evidence for Gamma-Ray Halos Around Active Galactic
  Nuclei and the First Measurement of Intergalactic Magnetic Fields},'' {\em
  Astrophys. J. Lett.} {\bf 722} (2010) L39,
  \href{https://arxiv.org/abs/1005.1924}{{\tt 1005.1924}}.

\bibitem{Tavecchio_2010}
F.~Tavecchio, G.~Ghisellini, L.~Foschini, G.~Bonnoli, G.~Ghirlanda, and
  P.~Coppi, ``The intergalactic magnetic field constrained by fermi/large area
  telescope observations of the {TeV} blazar 1es 0229 + 200,'' {\em Monthly
  Notices of the Royal Astronomical Society: Letters} (jun, 2010) no--no.

\bibitem{Neronov_2010}
A.~Neronov and I.~Vovk, ``Evidence for strong extragalactic magnetic fields
  from fermi observations of {TeV} blazars,'' {\em Science} {\bf 328} (apr,
  2010) 73--75.

\bibitem{Tavecchio_2011}
F.~Tavecchio, G.~Ghisellini, G.~Bonnoli, and L.~Foschini, ``Extreme {TeV}
  blazars and the intergalactic magnetic field,'' {\em Monthly Notices of the
  Royal Astronomical Society} {\bf 414} (apr, 2011) 3566--3576.

\bibitem{Essey:2010nd}
W.~Essey, S.~Ando, and A.~Kusenko, ``{Determination of intergalactic magnetic
  fields from gamma ray data},'' {\em Astropart. Phys.} {\bf 35} (2011)
  135--139, \href{https://arxiv.org/abs/1012.5313}{{\tt 1012.5313}}.

\bibitem{Finke:2015ona}
J.~D. Finke, L.~C. Reyes, M.~Georganopoulos, K.~Reynolds, M.~Ajello, S.~J.
  Fegan, and K.~McCann, ``{Constraints on the Intergalactic Magnetic Field with
  Gamma-Ray Observations of Blazars},'' {\em Astrophys. J.} {\bf 814} (2015),
  no.~1 20, \href{https://arxiv.org/abs/1510.02485}{{\tt 1510.02485}}.

\bibitem{PhysRevD.37.2743}
M.~S. Turner and L.~M. Widrow, ``Inflation-produced, large-scale magnetic
  fields,'' {\em Phys. Rev. D} {\bf 37} (May, 1988) 2743--2754.

\bibitem{1992ApJ...391L...1R}
B.~{Ratra}, ``{Cosmological ``Seed'' Magnetic Field from Inflation},'' {\em
  Astrophys. J. Lett.} {\bf 391} (May, 1992) L1.

\bibitem{PhysRevD.46.5346}
W.~D. Garretson, G.~B. Field, and S.~M. Carroll, ``Primordial magnetic fields
  from pseudo goldstone bosons,'' {\em Phys. Rev. D} {\bf 46} (Dec, 1992)
  5346--5351.

\bibitem{Dolgov:1993vg}
A.~Dolgov, ``{Breaking of conformal invariance and electromagnetic field
  generation in the universe},'' {\em Phys. Rev. D} {\bf 48} (1993) 2499--2501,
  \href{https://arxiv.org/abs/hep-ph/9301280}{{\tt hep-ph/9301280}}.

\bibitem{Gasperini:1995dh}
M.~Gasperini, M.~Giovannini, and G.~Veneziano, ``{Primordial magnetic fields
  from string cosmology},'' {\em Phys. Rev. Lett.} {\bf 75} (1995) 3796--3799,
  \href{https://arxiv.org/abs/hep-th/9504083}{{\tt hep-th/9504083}}.

\bibitem{Martin:2007ue}
J.~Martin and J.~Yokoyama, ``{Generation of Large-Scale Magnetic Fields in
  Single-Field Inflation},'' {\em JCAP} {\bf 01} (2008) 025,
  \href{https://arxiv.org/abs/0711.4307}{{\tt 0711.4307}}.

\bibitem{Demozzi:2009fu}
V.~Demozzi, V.~Mukhanov, and H.~Rubinstein, ``{Magnetic fields from
  inflation?},'' {\em JCAP} {\bf 08} (2009) 025,
  \href{https://arxiv.org/abs/0907.1030}{{\tt 0907.1030}}.

\bibitem{Kanno:2009ei}
S.~Kanno, J.~Soda, and M.-a. Watanabe, ``{Cosmological Magnetic Fields from
  Inflation and Backreaction},'' {\em JCAP} {\bf 12} (2009) 009,
  \href{https://arxiv.org/abs/0908.3509}{{\tt 0908.3509}}.

\bibitem{Emami:2009vd}
R.~Emami, H.~Firouzjahi, and M.~S. Movahed, ``{Inflation from Charged Scalar
  and Primordial Magnetic Fields?},'' {\em Phys. Rev. D} {\bf 81} (2010)
  083526, \href{https://arxiv.org/abs/0908.4161}{{\tt 0908.4161}}.

\bibitem{Bamba:2003av}
K.~Bamba and J.~Yokoyama, ``{Large scale magnetic fields from inflation in
  dilaton electromagnetism},'' {\em Phys. Rev. D} {\bf 69} (2004) 043507,
  \href{https://arxiv.org/abs/astro-ph/0310824}{{\tt astro-ph/0310824}}.

\bibitem{Bamba:2006ga}
K.~Bamba and M.~Sasaki, ``{Large-scale magnetic fields in the inflationary
  universe},'' {\em JCAP} {\bf 02} (2007) 030,
  \href{https://arxiv.org/abs/astro-ph/0611701}{{\tt astro-ph/0611701}}.

\bibitem{Kobayashi:2014sga}
T.~Kobayashi, ``{Primordial Magnetic Fields from the Post-Inflationary
  Universe},'' {\em JCAP} {\bf 05} (2014) 040,
  \href{https://arxiv.org/abs/1403.5168}{{\tt 1403.5168}}.

\bibitem{Barnaby:2012tk}
N.~Barnaby, R.~Namba, and M.~Peloso, ``{Observable non-gaussianity from gauge
  field production in slow roll inflation, and a challenging connection with
  magnetogenesis},'' {\em Phys. Rev. D} {\bf 85} (2012) 123523,
  \href{https://arxiv.org/abs/1202.1469}{{\tt 1202.1469}}.

\bibitem{BazrafshanMoghaddam:2017zgx}
H.~Bazrafshan~Moghaddam, E.~McDonough, R.~Namba, and R.~H. Brandenberger,
  ``{Inflationary magneto-(non)genesis, increasing kinetic couplings, and the
  strong coupling problem},'' {\em Class. Quant. Grav.} {\bf 35} (2018), no.~10
  105015, \href{https://arxiv.org/abs/1707.05820}{{\tt 1707.05820}}.

\bibitem{Durrer:2022emo}
R.~Durrer, O.~Sobol, and S.~Vilchinskii, ``{Magnetogenesis in Higgs-Starobinsky
  inflation},'' \href{https://arxiv.org/abs/2207.05030}{{\tt 2207.05030}}.

\bibitem{Kushwaha:2020nfa}
A.~Kushwaha and S.~Shankaranarayanan, ``{Helical magnetic fields from Riemann
  coupling},'' {\em Phys. Rev. D} {\bf 102} (2020), no.~10 103528,
  \href{https://arxiv.org/abs/2008.10825}{{\tt 2008.10825}}.

\bibitem{Kandus:2010nw}
A.~Kandus, K.~E. Kunze, and C.~G. Tsagas, ``{Primordial magnetogenesis},'' {\em
  Phys. Rept.} {\bf 505} (2011) 1--58,
  \href{https://arxiv.org/abs/1007.3891}{{\tt 1007.3891}}.

\bibitem{Subramanian:2009fu}
K.~Subramanian, ``{Magnetic fields in the early universe},'' {\em Astron.
  Nachr.} {\bf 331} (2010) 110--120,
  \href{https://arxiv.org/abs/0911.4771}{{\tt 0911.4771}}.

\bibitem{Mazzitelli:1995mp}
F.~D. Mazzitelli and F.~M. Spedalieri, ``{Scalar electrodynamics and primordial
  magnetic fields},'' {\em Phys. Rev. D} {\bf 52} (1995) 6694--6699,
  \href{https://arxiv.org/abs/astro-ph/9505140}{{\tt astro-ph/9505140}}.

\bibitem{Giovannini:2007rh}
M.~Giovannini, ``{Magnetogenesis, spectator fields and CMB signatures},'' {\em
  Phys. Lett. B} {\bf 659} (2008) 661--668,
  \href{https://arxiv.org/abs/0711.3273}{{\tt 0711.3273}}.

\bibitem{Patel:2019isj}
T.~Patel, H.~Tashiro, and Y.~Urakawa, ``{Resonant magnetogenesis from
  axions},'' {\em JCAP} {\bf 01} (2020) 043,
  \href{https://arxiv.org/abs/1909.00288}{{\tt 1909.00288}}.

\bibitem{Giovannini:2021thf}
M.~Giovannini, ``{Large-scale gauge spectra and pseudoscalar couplings},'' {\em
  Phys. Rev. D} {\bf 104} (2021), no.~12 123509,
  \href{https://arxiv.org/abs/2106.14927}{{\tt 2106.14927}}.

\bibitem{Talebian:2021dfq}
A.~Talebian, A.~Nassiri-Rad, and H.~Firouzjahi, ``{Primordial helical magnetic
  fields from inflation?},'' {\em Phys. Rev. D} {\bf 105} (2022), no.~2 023528,
  \href{https://arxiv.org/abs/2111.02147}{{\tt 2111.02147}}.

\bibitem{Talebian:2020drj}
A.~Talebian, A.~Nassiri-Rad, and H.~Firouzjahi, ``{Revisiting magnetogenesis
  during inflation},'' {\em Phys. Rev. D} {\bf 102} (2020), no.~10 103508,
  \href{https://arxiv.org/abs/2007.11066}{{\tt 2007.11066}}.

\bibitem{Fujita:2012rb}
T.~Fujita and S.~Mukohyama, ``{Universal upper limit on inflation energy scale
  from cosmic magnetic field},'' {\em JCAP} {\bf 10} (2012) 034,
  \href{https://arxiv.org/abs/1205.5031}{{\tt 1205.5031}}.

\bibitem{Ferreira:2013sqa}
R.~J.~Z. Ferreira, R.~K. Jain, and M.~S. Sloth, ``{Inflationary magnetogenesis
  without the strong coupling problem},'' {\em JCAP} {\bf 10} (2013) 004,
  \href{https://arxiv.org/abs/1305.7151}{{\tt 1305.7151}}.

\bibitem{Ferreira:2014hma}
R.~J.~Z. Ferreira, R.~K. Jain, and M.~S. Sloth, ``{Inflationary Magnetogenesis
  without the Strong Coupling Problem II: Constraints from CMB anisotropies and
  B-modes},'' {\em JCAP} {\bf 06} (2014) 053,
  \href{https://arxiv.org/abs/1403.5516}{{\tt 1403.5516}}.

\bibitem{Green:2015fss}
D.~Green and T.~Kobayashi, ``{Constraints on Primordial Magnetic Fields from
  Inflation},'' {\em JCAP} {\bf 03} (2016) 010,
  \href{https://arxiv.org/abs/1511.08793}{{\tt 1511.08793}}.

\bibitem{Fujita:2016qab}
T.~Fujita and R.~Namba, ``{Pre-reheating Magnetogenesis in the Kinetic Coupling
  Model},'' {\em Phys. Rev. D} {\bf 94} (2016), no.~4 043523,
  \href{https://arxiv.org/abs/1602.05673}{{\tt 1602.05673}}.

\bibitem{Giovannini:2013rme}
M.~Giovannini, ``{Fluctuations of inflationary magnetogenesis},'' {\em Phys.
  Rev. D} {\bf 87} (2013), no.~8 083004,
  \href{https://arxiv.org/abs/1302.2243}{{\tt 1302.2243}}.

\bibitem{Fujita:2013qxa}
T.~Fujita and S.~Yokoyama, ``{Higher order statistics of curvature
  perturbations in IFF model and its Planck constraints},'' {\em JCAP} {\bf 09}
  (2013) 009, \href{https://arxiv.org/abs/1306.2992}{{\tt 1306.2992}}.

\bibitem{Byrnes:2011aa}
C.~T. Byrnes, L.~Hollenstein, R.~K. Jain, and F.~R. Urban, ``{Resonant magnetic
  fields from inflation},'' {\em JCAP} {\bf 03} (2012) 009,
  \href{https://arxiv.org/abs/1111.2030}{{\tt 1111.2030}}.

\bibitem{Turner:1987bw}
M.~S. Turner and L.~M. Widrow, ``{Inflation Produced, Large Scale Magnetic
  Fields},'' {\em Phys. Rev. D} {\bf 37} (1988) 2743.

\bibitem{whittaker_watson_1996}
E.~T. Whittaker and G.~N. Watson, {\em A Course of Modern Analysis}.
\newblock Cambridge Mathematical Library. Cambridge University Press, 4~ed.,
  1996.

\bibitem{Enqvist:2016mqj}
K.~Enqvist, M.~Karciauskas, O.~Lebedev, S.~Rusak, and M.~Zatta,
  ``{Postinflationary vacuum instability and Higgs-inflaton couplings},'' {\em
  JCAP} {\bf 11} (2016) 025, \href{https://arxiv.org/abs/1608.08848}{{\tt
  1608.08848}}.

\bibitem{Lachapelle:2008sy}
J.~Lachapelle and R.~H. Brandenberger, ``{Preheating with Non-Standard Kinetic
  Term},'' {\em JCAP} {\bf 04} (2009) 020,
  \href{https://arxiv.org/abs/0808.0936}{{\tt 0808.0936}}.

\bibitem{Shtanov:2020gjp}
Y.~Shtanov and M.~Pavliuk, ``{Model-independent constraints in inflationary
  magnetogenesis},'' {\em JCAP} {\bf 08} (2020) 042,
  \href{https://arxiv.org/abs/2004.00947}{{\tt 2004.00947}}.

\bibitem{McLachlanN.W.NormanWilliam1964Taao}
N.~W. N.~W. McLachlan, {\em Theory and application of Mathieu functions / by
  N.W. McLachlan.}
\newblock Dover books on engineering and engineering physics. Dover
  Publications, New York, 1964.

\bibitem{Akrami:2017cir}
Y.~Akrami, R.~Kallosh, A.~Linde, and V.~Vardanyan, ``{Dark energy,
  $\alpha$-attractors, and large-scale structure surveys},'' {\em JCAP} {\bf
  06} (2018) 041, \href{https://arxiv.org/abs/1712.09693}{{\tt 1712.09693}}.

\bibitem{Rubio:2017gty}
J.~Rubio and C.~Wetterich, ``{Emergent scale symmetry: Connecting inflation and
  dark energy},'' {\em Phys. Rev. D} {\bf 96} (2017), no.~6 063509,
  \href{https://arxiv.org/abs/1705.00552}{{\tt 1705.00552}}.

\bibitem{Schwinger:1951nm}
J.~S. Schwinger, ``{On gauge invariance and vacuum polarization},'' {\em Phys.
  Rev.} {\bf 82} (1951) 664--679.

\bibitem{Dunne:1998ni}
G.~V. Dunne and T.~Hall, ``{On the QED effective action in time dependent
  electric backgrounds},'' {\em Phys. Rev. D} {\bf 58} (1998) 105022,
  \href{https://arxiv.org/abs/hep-th/9807031}{{\tt hep-th/9807031}}.

\bibitem{Chu:2010xc}
Y.-Z. Chu and T.~Vachaspati, ``{Capacitor Discharge and Vacuum Resistance in
  Massless QED$_{2}$},'' {\em Phys. Rev. D} {\bf 81} (2010) 085020,
  \href{https://arxiv.org/abs/1001.2559}{{\tt 1001.2559}}.

\bibitem{Gold:2020qzr}
G.~Gold, D.~A. Mcgady, S.~P. Patil, and V.~Vardanyan, ``{Backreaction of
  Schwinger pair creation in massive QED$_{2}$},'' {\em JHEP} {\bf 10} (2021)
  072, \href{https://arxiv.org/abs/2012.15824}{{\tt 2012.15824}}.

\bibitem{Cai:2021yvq}
Y.-F. Cai, J.~Jiang, M.~Sasaki, V.~Vardanyan, and Z.~Zhou, ``{Beating the Lyth
  Bound by Parametric Resonance during Inflation},'' {\em Phys. Rev. Lett.}
  {\bf 127} (2021), no.~25 251301, \href{https://arxiv.org/abs/2105.12554}{{\tt
  2105.12554}}.

\bibitem{SALKUYEH2006442}
D.~K. Salkuyeh, ``Comments on “a note on a three-term recurrence for a
  tridiagonal matrix”,'' {\em Applied Mathematics and Computation} {\bf 176}
  (2006), no.~2 442--444.

\end{thebibliography}\endgroup

\end{document}